\documentclass{aa} % for a referee version
\usepackage{txfonts} 
\usepackage{graphicx} 
\usepackage{appendix} 
\usepackage{natbib} 
\usepackage{url} 
\usepackage{subfigure} 
\bibpunct{(}{)}{;}{a}{}{,} 
\newcommand{\kms}{\mbox{\,km\,s$^{-1}$}} 
 
\begin{document} 
 
\title{A revision of the X-ray absorption nature of BALQSOs \thanks{Based on observations obtained with {XMM-{\textit{Newton}}}, an ESA science mission with instruments and contributions directly funded by ESA Member States and NASA.}} 
 
\author{ 
	A.~Streblyanska\inst{1}, 
	X.~Barcons\inst{1}, 
	F. J.~Carrera\inst{1} 
	\and R. Gil-Merino\inst{1}} 
 
\offprints{A.~Streblyanska, \email{alina@ifca.unican.es }} 
 
\institute{\inst{1}Instituto de F\'{\i}sica de Cantabria (CSIC-UC), 39005 Santander, Spain } 
 
\date{Received 27 November 2009; accepted 22 January 2010} 
 
\abstract 
% context heading (optional - leave it empty if necessary) 
{}
% aims 
 {  Broad absorption line quasars (BALQSOs) are key objects for studying the structure and emission/absorption properties of AGN. However, despite their fundamental importance, the properties  of BALQSOs remain poorly understood. To investigate the X-ray nature of these sources, as well as the correlations between X-ray and rest-frame UV properties, we compile a large sample of BALQSOs observed by XMM-Newton. }
% methods
 {We collect  information for 88 sources from the literature and existing catalogues, creating the largest BALQSO sample analysed optically and in X-ray to date. We performed a full X-ray spectral analysis (using unabsorbed and both neutral and ionized absorption models) on a sample of 39 sources with higher X-ray spectral quality, and an approximate hardness-ratio analysis on the remaining sources. Using available optical spectra, we calculate the BALnicity index and investigate the dependence of this optical parameter on different X-ray properties.}
% results. 
 { Using the  neutral absorption model, we find that 36\% of our BALQSOs have $N_{\rm{H}}^{n} < 5\times10^{21}\:$cm$^{-2}$, lower than the expected X-ray absorption for these objects. However, when we used a physically-motivated model for the X-ray absorption in BALQSOs, i.e., ionized absorption, $\sim$90\% of the objects are absorbed. The observed difference in ionized properties of sources with the BALnicity index (BI)=0 and BI$>$0 might be explained by different physical conditions of the outflow and/or inclination effects. 
The absorption properties also suggest that LoBALs may be physically different objects from HiBALs. In addition, we report on a correlation between the  ionized absorption column density and BAL parameters. There is evidence (at the 98\% level) that the amount of X-ray absorption $is$ correlated with the strength of high-ionization UV absorption. Not previously reported, this correlation  can be naturally understood in virtually all BALQSO models, as being driven by the total amount of gas mass flowing towards the observer. We also find a hint of a correlation between the BI and the ionization level detected in X-rays.  }
% conclusions heading
{}

 
\titlerunning{BALQSOs} 
\authorrunning{A. Streblyanska et al.} 
 
\maketitle

\section{Introduction} 
 
The nature of intrinsic absorption in quasars has important implications for physical models of the AGN ``central engine'' and, in general, for developing the unified model. 
The subclass of broad absorption line (BAL) quasars, characterized by strong, broad, and blueshifted absorption troughs in their spectra, is the main manifestation of the importance of outflows in the AGN phenomenon.  While BAL objects are important in understanding the properties of quasars, the nature of BALQSOs remains poorly understood and there is still no consensus about whether BALQSOs differ intrinsically from non-BAL quasars.  The small percentage of BALQSOs among quasars ($\sim 15\%$) is generally interpreted as an orientation effect in the unified model \citep[QSOs viewed at large inclination angles close to their equatorial plane,][]{1991ApJ...373...23W}. However, the properties of some low-ionization BALQSOs appear inconsistent with simple unified models and can be explained only by an evolutionary scenario, where BALQSOs are young or recently refueled quasars \citep[e.g.,][]{1992ApJ...397..442B, 2000ApJ...538...72B}.  In this model, broad absorption lines appear when a nucleus blows off gas and dust during a dust-enshrouded quasar phase, evolving to non-BAL quasars. Some results for the radio, such as for radio BALQSOs that share several properties with young radio sources \citep[e.g.,][]{2008MNRAS.391..246L} and infrared, such as the 2MASS Infrared survey that contains an unusually high fraction of bright BALQSOs at high z \citep[see][]{2009ApJ...698.1095U} seem to be in favor of this idea.

Thus, the study of BALQSOs advances significantly  not only our understanding of the structure and emission/absorption physics of AGN, but also helps in developing an evolutionary scenario of quasars.

While the UV and optical properties of BALQSOs are more or less understandable and fit in with our ideas about the expected properties of this kind of objects, the X-ray properties of BALs remain controversial. Theoretical modelling of BALs \citep[e.g., the radiatively driven disk-wind paradigm by][]{1995ApJ...451..498M, 2000ApJ...543..686P} require X-ray column densities around $10^{22}-10^{23}$~cm$^{-2}$  to prevent over-ionization of the wind by the soft X-rays generated near the central engine. Until recently, the number of X-ray analysed BALQSOs was rather small and, in many cases, the optically selected bright objects were undetectable in X-rays, implying high column densities of absorbing gas and, therefore, supporting theoretical predictions. The hardness ratio analysis of 35 X-ray detected BALQSOs by \citet[][]{2006ApJ...644..709G}, also detects significant intrinsic absorption ($N_{\rm{H}}\sim 10^{22}-10^{24}$~cm$^{-2}$). However, a sample obtained by cross-correlating the Sloan Digital Sky Survey and the Second XMM-Newton Serendipitous Source (2XMM) catalogues by \citet{2008A&A...491..425G} infers lower values of neutral absorption than found in optically selected BALQSOs samples. While this result is expected, given that we require that the sources be not just detected in X-rays, but also have enough counts to perform X-ray spectral analysis, questions about the real fraction of the absorbed BALQSOs and properties of the absorbed gas remain open. Moreover, if we know that BALQSOs $are$ objects with complex absorbers, outflows and ionized material \citep[as implied by the definition and confirmed by observations of a few bright sources, e.g.,][]{2003AJ....126.1159G}, how correct is the application of the simple neutral absorbed model and how might this influence our view of BALQSOs?

Another important issue today is the definition of BALQSOs itself, which is a bit diffuse. The traditional BALs are defined to have CIV absorption troughs broader than 2000 \kms, this width ensuring that the absorption is from a nuclear outflow and effectively excluding associated absorbers. However, it could potentially exclude unusual or interesting BALs \citep[see][]{2000ApJ...538...72B,2002ApJS..141..267H}. Therefore, \citet{2006ApJS..165....1T}, in their SDSS sample, also included in the BALQSO class BAL absorption features at lower outflow velocities (within 3000 \kms). As a result, a significantly higher fraction of QSOs can be classified as BALQSOs \citep[e.g., in the 2MASS survey, the fraction of BALQSOs increases by a factor of two with the new definition,][]{2008ApJ...672..108D} and BALQSOs are objects with both weaker and much narrower absorption troughs than previously assumed. Despite the obvious weak points of this approach \citep[see][]{2008MNRAS.386.1426K}, this classification is widely used nowadays. Thus, an important question arises: is there any difference in physical properties between the ``classical'' and ``new'' BALQSOs?

To explore all the questions mentioned above, we created the most complete sample of X-ray detected BALQSOs to date, based on XMM-Newton observations. 
 
This paper is organized as follows. We present the sample in Section 2. In Section 3, we describe the observations and their reduction; the X-ray and optical data analyses are presented in Section 4. Section 5 shows the results obtained from the X-ray spectral and hardness ratio analyses. In Section 6, we discuss our results and implications for different models. 
%A summary of our main results is given in Sect. 7. 

A concordance cosmology model with $H_0=70$~\kms~Mpc$^{-1}$, $\Omega_{\Lambda}=0.7$, and $\Omega_{\rm{M}}=0.3$ is used throughout the paper.

\section{The sample}\label{sec:sample} 
 
 Based on the material producing the BAL troughs, the BALQSOs are classified into high-ionization BALQSOs (HiBALs) and low-ionization BALQSOs (LoBALs). HiBALs contain strong, deep, and broad absorption troughs shortward of high-ionization emission lines and are typically identified by the presence of C~IV$\lambda 1549$ absorption troughs.  In addition to HiBAL features, LoBALs show broad absorption troughs in the low-ionization species, such as  Mg~II$\lambda 2798$ and Al~III$\lambda 1857$ lines.  The rare subclass of LoBALs  ($\sim 1\%$) exhibiting broad absorption in metastable Fe~II and Fe~III lines is called FeLoBAL. About 10 to 20\% of all quasars are BALQSOs. Out of these, about 15$\%$ are LoBALs.

Traditionally, BALQSOs are characterized by their BALnicity index \citep[BI,][]{1991ApJ...373...23W}, which is a modified velocity equivalent width of the C~IV. The ``traditional'' BALQSOs have BI$\,>0$\footnote{That is, at least 2000 \kms\ wide C~IV absorption trough, blueshifted by $> 3000$ km s$^{-1}$ with respect to the  C~IV line. This definition is formally detailed in Sect.~\ref{sec:optic}.}. The ``extended'' definition uses the absorption index \citep[AI,][]{2002ApJS..141..267H,2006ApJS..165....1T} and includes BALQSOs that have absorption at outflow velocities within 3000 \kms\ of the emission-line redshift. We used both kinds of objects to create our sample.

%BI is computed considering absorption trough spanning $\geq 2000$ km s$^{-1}$ in width, absorbing at least 10\% of the continuum, and blueshifted by $> 3000$ km s$^{-1}$ with respect to the  C~IV line.  
 
%For these reasons, a less restrictive criterion to identify BALQSOs has been introduced in the last years, the absorption index \citep[AI,][]{2002ApJS..141..267H,2006ApJS..165....1T}. This index is computed in the same way as BI, but relaxing the 3000 km s$^{-1}$ blueshift criterion and considering all absorption troughs with a blueshift $> 0$~km s$^{-1}$ with respect to the corresponding emission lines, and with a width of at least 1000 \kms.  \\ 

We cross-correlate the 2XMM catalogue with the NASA Extragalactic Database 
(NED)\footnote{The NASA/IPAC Extragalactic Database (NED) is operated by the Jet 
Propulsion Laboratory, California Institute of Technology, under contract with the National Aeronautics and Space Administration.}, the largest extragalactic 
sources database. We preselected objects with Galactic latitude greater than 20 degrees and high quality X-ray  data corresponding to flag zero in the 2XMM  catalogue. The queries to NED were launched in two different modes: either directly as a socket query or with a batch file. The condition for source matching was to fall within a distance of less than 5~arcsecs or 5 times the error in the position of the X-ray source queried. We finally stored only sources of known redshift. The detailed description of the cross-correlation and algorithm used is discussed in Gil-Merino et al. (2010, in preparation).

In this way, we include SDSS BALQSOs by \citet{2006ApJS..165....1T} and sources from individual random observations. In addition, we cross-correlate the 2XMM catalogue with other BALQSOs samples \citep[][]{2008ApJ...680..169S,2009ApJ...692..758G} based on the SDSS DR5 catalogue \citep {2007AJ....134..102S}.  These recently published catalogues have not yet been included in NED and, therefore, are missed in our main cross-correlation procedure.    
 
%\citet{2008ApJ...680..169S} (the sources were selected using the classical BI $>$ 0 criterium) and \citet{2009ApJ...692..758G} \citep[new extended version of the catalog by][]{2006ApJS..165....1T}. While the most of the their sources coinsiden with sources from \citet{2006ApJS..165....1T},  we found new matched objects (8 from  \citet{2009ApJ...692..758G} and 21 from \citet{2008ApJ...680..169S}).our analysis (54 sources from Giustini et al. 2008), and more correct than Fan et al. 2009, where 17 sources out of 41 analysed sources have not corresponded X-ray sources. 
 
\begin{figure} 
\centering 
\resizebox{\hsize}{!}{\includegraphics{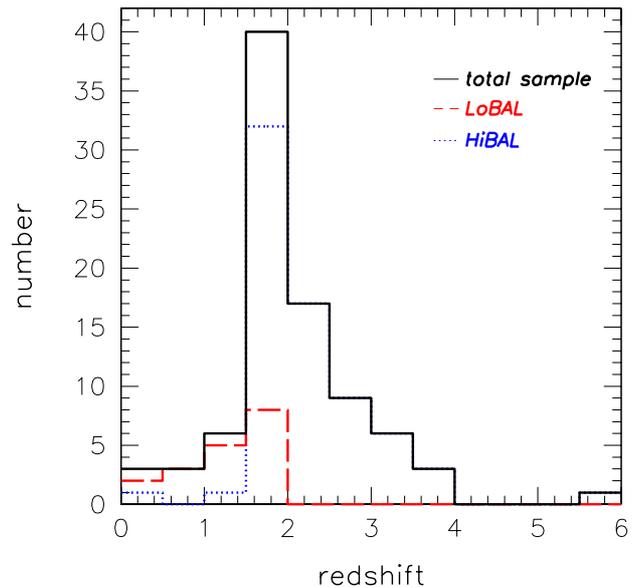}} 
\caption{\label{fig:zdis} Redshift distribution of the BALQSOs in our sample. The dotted and dashed lines correspond to the HiBALs and LoBALs, respectively. The upper solid line shows the distribution for the total sample of 88 sources. Most of the sources of both types have z around 1.9, and constitute 45\% of all the sources.}  
\end{figure} 
 
Our final sample consists of 88 sources (70 HiBALs and 18 LoBALs) and spans the redshift range $0.15 < z < 5.8$ with a peak at $z \sim 1.9$ (see Fig.~\ref{fig:zdis}). Among the selected BALQSOs, we performed a full X-ray spectral analysis of 39 sources and an X-ray hardness ratio spectral analysis of the remaining 49 sources.

The details of each source in our sample are presented in Table \ref{table:1}.  
 The first and second columns list the NED and 2XMM names, respectively. In the next columns, the redshift, the BAL type, the value of BALnicity Index (see details in Sect.~\ref{sec:optic}), an abbreviation for the performed X-ray analysis,  and a number of the individual XMM-Newton observations are reported. The last column shows the notes on individual sources, i.e., details about the calculation of the BI or the reference from which we gathered information about the object.

\section{Observations and data reduction} 
\subsection{X-ray data} 
 
For the full spectral analysis, we selected sources with $\gtrsim 70$ counts detected by the EPIC-pn instrument in the $0.2-10$~keV energy band. For each observation, the X-ray data  were retrieved from the {\textit{XMM-Newton}} Science Archive (XSA) and reduced with {\textit{SAS}} v8.0.0 and the latest calibration files. The background light curves at energies above 10 keV were used to filter the data and remove high background flaring periods. We used the \texttt{eregionanalyse} task for source region optimization and maximization of the signal-to-noise ratio. The \texttt{evselect} tool was used to extract the spectrum and background region, which was defined as a circle around the source, after masking out nearby objects. We extracted the spectra of each source for the pn, MOS1, and MOS2 detectors using only events with pattern 0--4 (single and double) for the pn and 0--12 for the MOS cameras.  All spectra were extracted in the $0.2-10$~keV band, EPIC being the most accurately calibrated. For each spectrum, we generated  a redistribution matrix file (RMF) and ancillary response file (ARF) using the \texttt{rmfgen} and \texttt{arfgen} tasks, respectively.  
 
To improve statistics, MOS1 and MOS2 source spectra were combined into a mean MOS spectrum by summing the counts from the channels with the nominal energy range using our own \texttt{perl} script. The background spectra and calibration files were merged in the same fashion. We combined the spectra by weighting them for the exposure time of the individual detectors and taking into account the value of the \texttt{backscale} parameter. If the source was observed more than once at compatible off-axis angles, we merged the pn, mean MOS spectra, and the calibration files from one observation with the corresponding spectra and files from other observations. This approach allowed us to measure the time-averaged spectral properties (i.e., the flux and spectral shape), ignoring possible time variability of the sources.

\subsection{Optical data} 
 
For sources with SDSS counterparts (69 out of 88), we retrieved optical spectra from the SDSS database\footnote{http://cas.sdss.org/astro/en/tools/search/SQS.asp}. Spectra for the remaining 19 objects were either  provided by authors or reconstructed, using a special tool, from scanned graphs of the corresponding articles (see details in the last column of Table \ref{table:1}).

\section{X-ray and optical data analysis} 
\subsection{X-ray Analysis} 
 
A full spectral analysis was performed for a sample of 39 sources with the highest quality statistics, leading to proper values of $\Gamma$ and $N_{\rm{H}}$ in the source rest-frame. We used XSPEC v12.4.0 \citep{1996ASPC..101...17A} to perform the spectral analysis. All spectra were binned to a minimum of 3 counts/bin. We performed this minimal grouping to avoid channels with no counts, i.e., the spectrum is essentially unbinned and no spectral information is lost.  Because of the small number of counts in our spectra, we decided to use the Cash-statistic \citep[XSPEC \texttt{C-stat,}][]{1979ApJ...228..939C}. The new version of XSPEC allows C-statistic fits to data for which background spectra are considered. Comparison of the fitting results from Cash and $\chi^2$ statistics (for our best quality spectra) reveals that the difference is negligible.  Since C-stat does not indicate at all of quality of the fit, we computed its goodness using Monte Carlo probability calculations.  For this purpose, we used the \texttt{goodness} command in XSPEC, which performs Monte Carlo simulations of 100 model spectra using the best-fit  model and infers the percentage of simulated spectra that had a fit statistic less than that obtained from the fit to the real data. The `goodness of fit' should be around 50\%, if the observed data were produced by the fitted model. The obtained values were, indeed, around this value for all our spectra. 
%All quoted errors are at 90\% confidence level for one parameter of interest \citep[i.e. $\Delta\chi^2=2.706$,][]{1976ApJ...210..642A} except otherwise stated.  

 EPIC-pn and MOS data were fitted simultaneously using appropriate models. We, initially, fit the spectra with a model consisting of a power law with an intrinsic absorption component at the source redshift \texttt{zphabs}, and an additional photoelectric absorption component \texttt{phabs} that was fixed at the Galactic column density \citep{1990ARA&A..28..215D}. From our model fits, we computed the slope of a power-law spectrum (photon index $\Gamma$), the intrinsic rest-frame neutral column density  $N_{\rm{H}}^{n}$, and the X-ray flux and luminosity in the $0.5-2$ and $2-10$ keV bands. For absorbed sources, the rest-frame luminosities were corrected for absorption by setting to zero the $N_{\rm{H}}$ values in the XSPEC best-fit model. Spectral analysis results are reported in Table \ref{table:2}.  
 
Fourteen of the 39 BALQSOs ($\sim$36\%) have low intrinsic neutral absorption, $N_{\rm{H}}^{n}<5\times 10^{21}\:$cm$^{-2}$, confirming the result ($\sim$36\%) of \citet{2008A&A...491..425G} . 
 
\begin{figure} 
\centering 
\resizebox{\hsize}{!}{\includegraphics{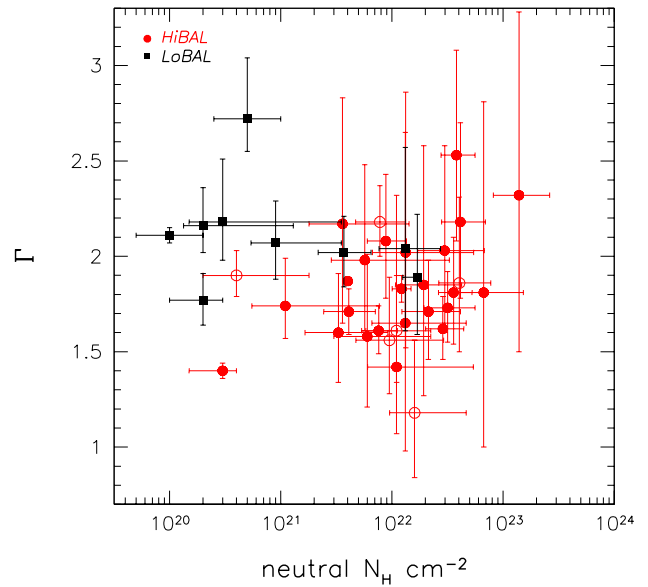}} 
\caption{\label{fig:gam-nh}. The power-law photon index $\Gamma$ versus rest-frame $N_{\rm{H}}^{n}$ for the analysed  sample of 39 sources suitable for full spectral analysis. Solid black squares and red circles refer to LoBALs and HiBALs, respectively. Open red circles indicate HiBALs with BI$=$0. The parameter $\Gamma$ for our objects ranges from 1.18 to 2.72, the majority of the sources clustering around 1.9.}  
\end{figure} 

As noted in the Introduction, this result is difficult to understand in the context of any BALQSO scenario.  For this reason, we decided to study the possibility that ionized (as opposed to neutral) gas is significantly present in our BALQSOs, and then characterise the properties of this ionized absorber. If BALQSOs $are$  special sources with strong outflowing ionized winds and ionized or partially covering intrinsic absorption, then we need to adopt a model that takes account of the presence of these features.   To test this, we replaced the neutral absorber in our spectral model with an ionized intrinsic absorber, modelled using the XSPEC model \texttt{absori} \citep{1992ApJ...395..275D}. Owing to the low signal-to-noise ratio of the data, we forced the \texttt{absori} photon index to be fixed and equal to the continuum power-law index corresponding to $\Gamma$ obtained from the previous absorbed power-law model. The temperature of the absorbing material was fixed at $T=3.0 \times 10^{4}\;\rm K$ and its redshift was fixed to be the redshift of the source.  
%In order to consider the possible effects of ionization of the absorbing medium in our sources, we employed the absori model. This is a simple single-zone photoionization model which assumes a power-law input spectrum
 
To constrain the output parameters of the ionized hydrogen column density $N_{\rm{H}}^{i}$ and ionization parameter $\xi$\footnote{$\xi=L/nr^{2}$, where $L$ is the ionizing luminosity of the source, $n$ is the number density of the absorber, and $r$ is the distance between the absorber and the ionizing source}, we varied the values of the parameter $\xi$ from 10 to 1000 erg cm s$^{-1}$ in steps of 50 erg cm s$^{-1}$\footnote{We choose this range of $\xi$ values in order to reproduce possible ionization states of the material, i.e., from almost neutral ($\xi<$ 50 erg cm s$^{-1}$) to highly ionized ($\xi\sim$ 1000 erg cm s$^{-1}$) absorption} to find best-fit values of $N_{\rm{H}}^{i}$, and then we used these values of $N_{\rm{H}}^{i}$ to constrain $\xi$. Each of these steps was checked against the `goodness of fit' by the simulation of 100 model spectra. Using this approach, we found border values of two parameters for each of our sources.  Almost as expected, this model produces higher values of absorption for all of our sources (only 3 sources out of 39 still have $N_{\rm{H}}^{i}<5\times 10^{21}\:$cm$^{-2}$). This is consistent with a scenario un which there is a significant amount of gas along our line of sight to the nuclei of BALQSOs, but this gas is likely to be ionized in most cases. The estimated parameters for the \texttt{absori} model are listed in Table~\ref{table:3}. 
 
%All model parameters are kept linked in the case of the two MOS. we varied for each source $\Gamma$ and $NH$ around their best-fit values. 
%We used the XSPEC model \texttt{ABSORI} \citep{1992ApJ...395..275D} to model the intrinsic absorption component as an ionized absorber. xi=ionization level of absorbing material,, absorber ionization state. 
% we added another multiplicative component. we  fitted the absorption through ionized material as represented by the XSPEC model \texttt{ABSORI} 

\begin{figure} 
\centering 
\includegraphics[width=8.5cm]{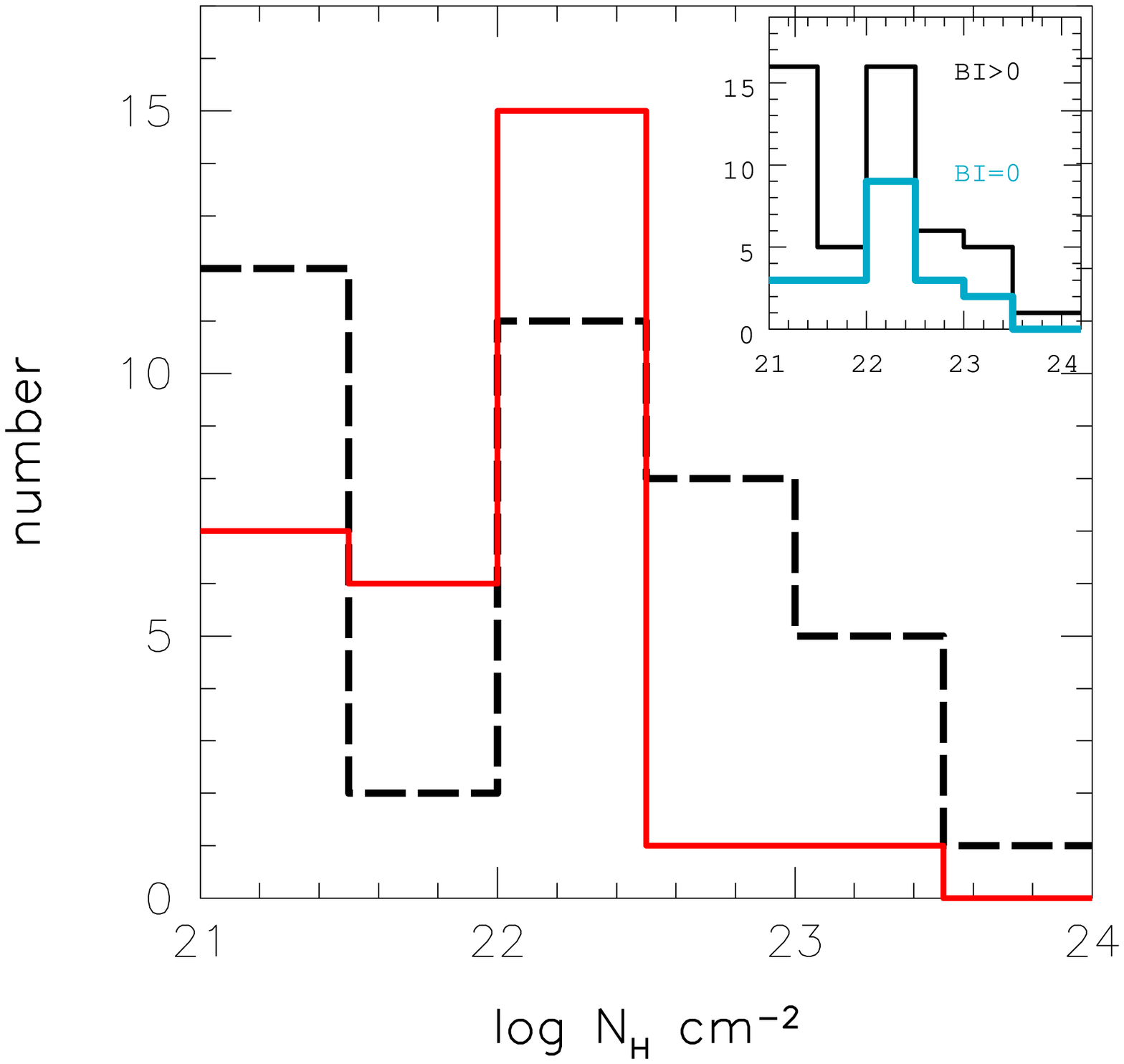} 
\includegraphics[width=8.5cm]{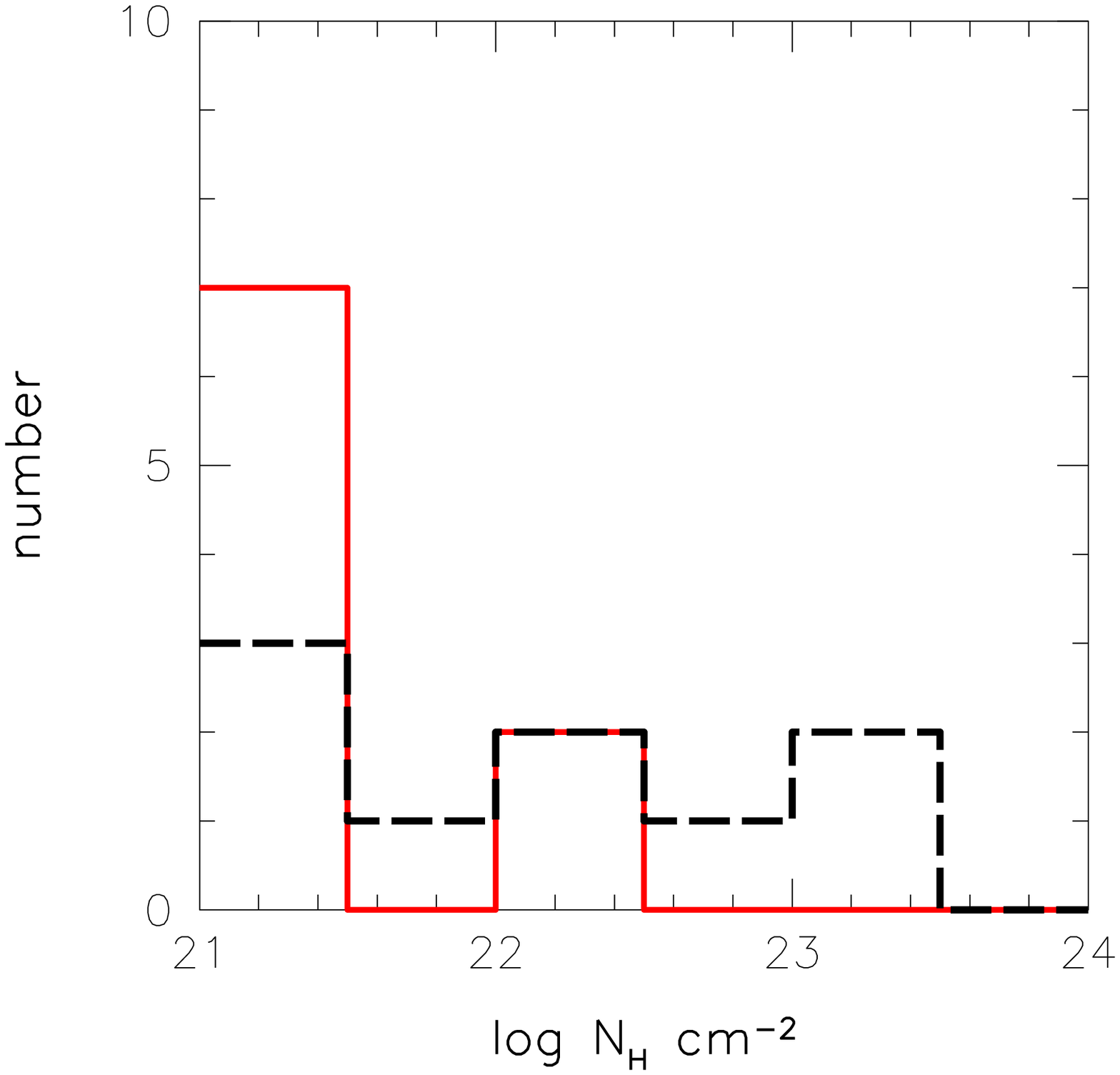} 
\caption{\label{fig:nh-numbers} Top panel: Distribution of neutral absorbing column densities for 70 HiBALs (top panel) and 18 LoBALs (bottom panel) in our sample.  In both graphics, the red solid lines refer to values obtained from the full spectral analysis of our high quality spectra (30 HiBALs and 9 LoBALs, respectively), while black dashed lines refer to values obtained from the HR spectral analysis of the remaining sources. The insert in the top panel shows the observed distribution of absorption for the total sample of 70 HiBALs divided by values of the BALnicity Index. The thin black line refers to the 49 sources with BI$>$0 and thick blue line refers to the 21 sources with BI$=$0. }    
\end{figure}

We performed a hardness ratio (HR) spectral analysis of our lower spectral quality sources (49 out of 88). The hardness ratio was calculated using the standard formula ${\rm HR} =(H - S)/(H + S)$, where $H$ and $S$ correspond to counts in the hard ($2.0-10.0$~keV) and soft ($0.5-2.0$~keV) energy bands, respectively.  Using XSPEC, we calculated the HRs for a grid of power-law models with the canonical value of $\Gamma=1.8$ and neutral absorption in the range of 10$^{21}$ -- 10$^{24}$ cm$^{-2}$  at the source redshift. Both the Galactic and intrinsic absorption were included in the model, details of the calculation being present in  \citet{2009}. We then compared the observed hardness ratio of each source with the modelled ones, and derived the amount of the intrinsic absorption. The absorbing column densities and their 1$\sigma$ errors (propagated from count-rate errors) are reported in Table~\ref{table:4}. In Fig.~\ref{fig:nh-numbers}, we  show the absorption distributions for our HiBAL (top panel) and LoBAL (bottom panel) subsamples. For each subsample, we plot the absorption values obtained from the HR analysis of our lower spectral quality sources and the fitted results for the 39 spectrally analysed sources.

\subsection{Optical data analysis}\label{sec:optic} 
 
 Following the definition given by \citet{1991ApJ...373...23W}, we calculated the BALnicity index of C IV to be:  

\begin{equation} 
BI=\int^{25000}_{3000}\left [1-\frac{f(v)}{0.9} \right] C dv , 
\end{equation} 
where $f(v)$ is the normalized flux (calculated from observed and fitted fluxes) as a function of velocity in km s$^{-1}$, and $C = 1$ at trough velocities greater than 2000 \kms\ from the start of a contiguous trough, and $C = 0$ elsewhere. In the case of the calculation of BI from Mg II, which is generally narrower than the high-ionization lines, the integral in Eq. (1) starts from $v=0$ \kms\ and $C = 1$ at trough velocities more than 1000 \kms.  The BALnicity index can take on any value in the range 0~\kms$\leqslant$~BI~$\leqslant$20,000~\kms.

Because of the redshift range of our LoBALs, only for 3 sources do we have BI measurements from C IV, and for the remaining objects (at z $<1.8$) the  Mg II BI is presented (see Table~\ref{table:1}). 
 
%BI a sum of the "modified equivalent width" of the portions of all contiguous BAL troughs between 3000 and 25,000 \kms shortward of 1549 \AA. The equivalent width is "modified" because only those parts of the troughs beyond the first 2000 \kms of width that dip below 10\% of the continuum are included.  
%We adopted the conventional definition for the BAL QSO: the equivalent width (in km s-1 ) of any contiguous absorption (at least 10% below the continuum) that falls between 3000-25,000 km s-1 , blueshifted from the systematic redshift, exceeds 2000 km s-1  
%In the traditional definition by \citet{1991ApJ...373...23W} it state that BALs must be at least 2000 \kms wide and at least 10\% below the continuum at maximum depth.  

 \begin{figure} 
\centering 
\includegraphics[width=8.5cm]{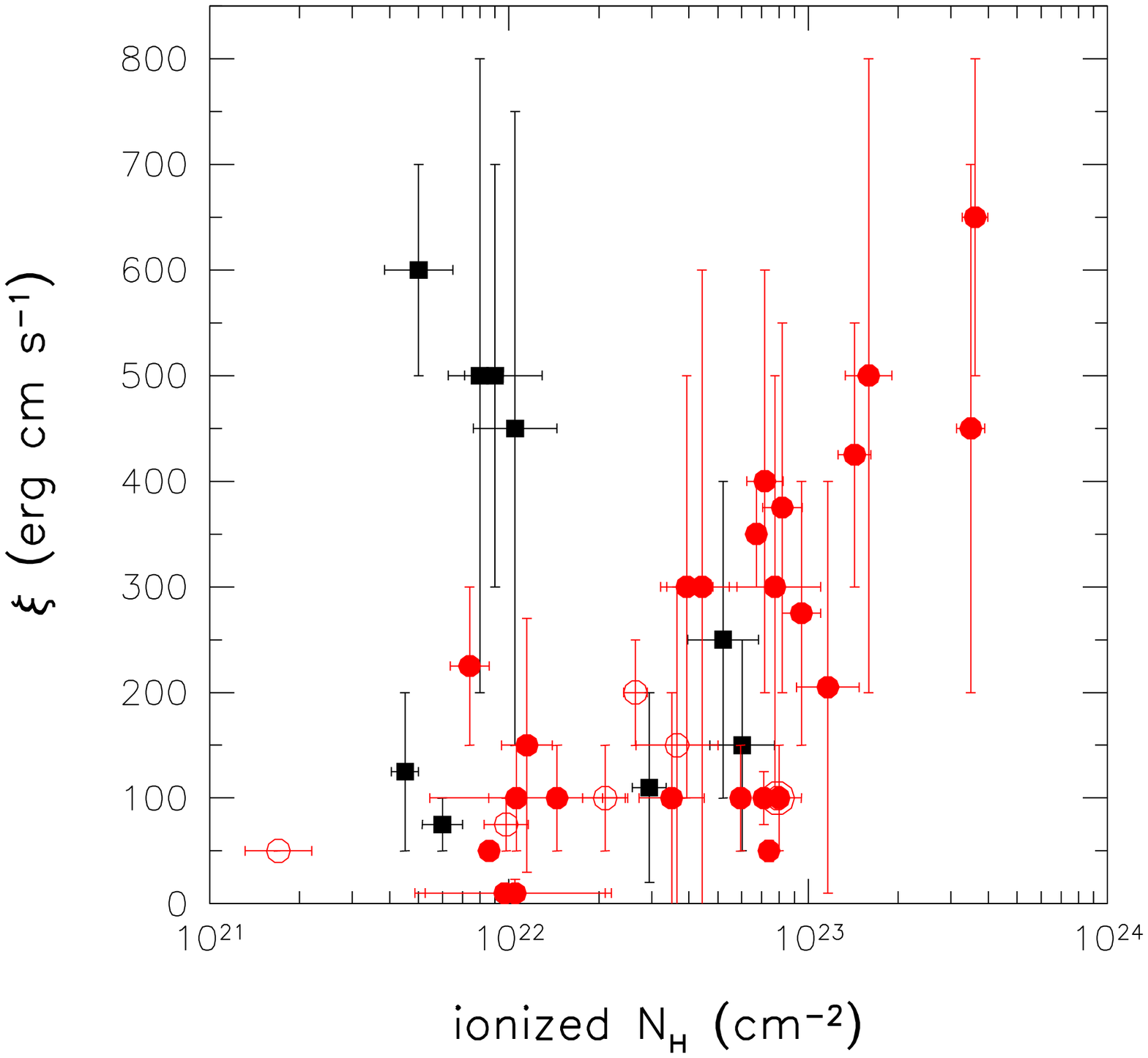} 
\includegraphics[width=8.5cm]{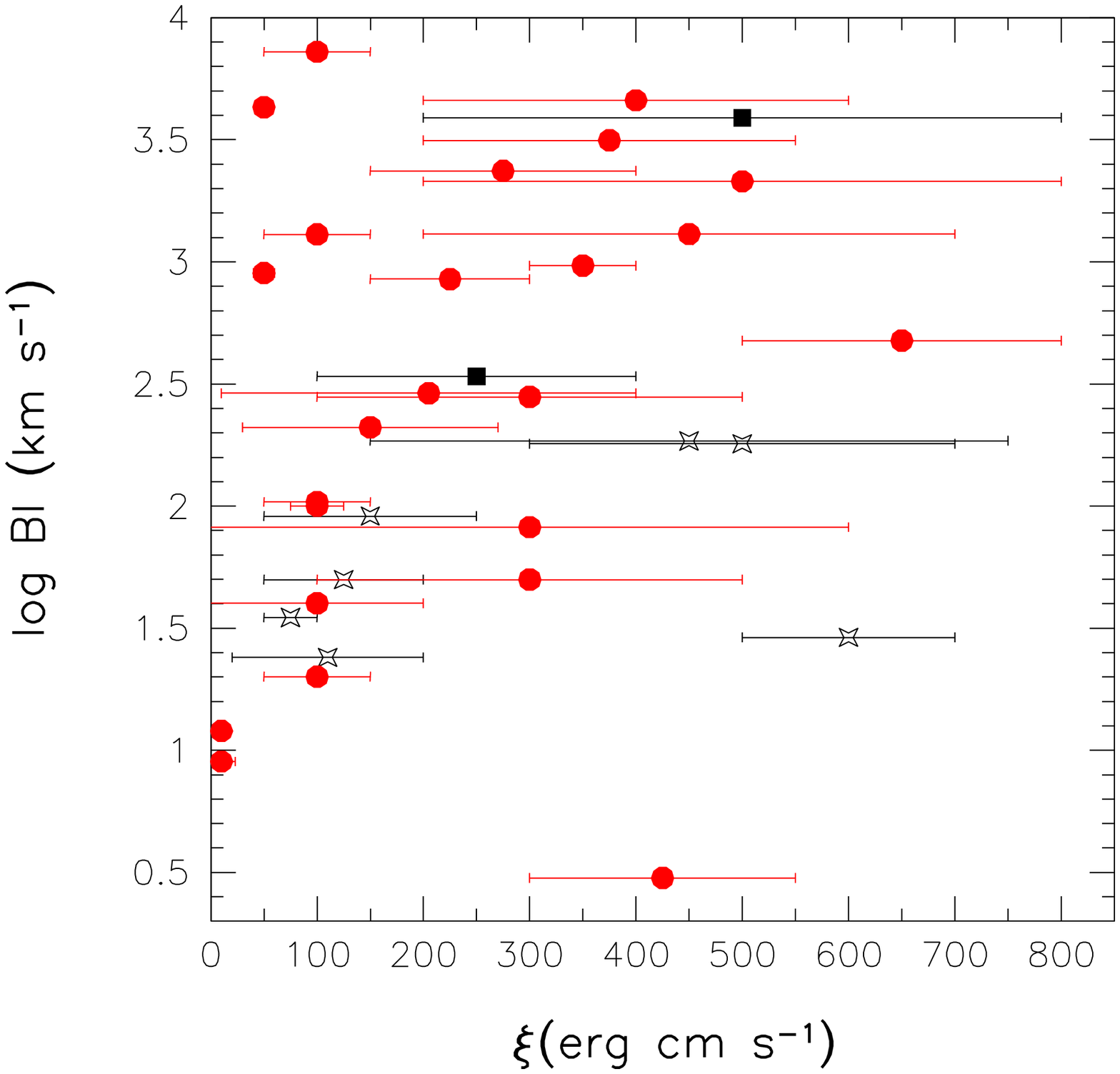} 
\caption{\label{fig:nh} Top panel: The ionization level of absorbing material $\xi$ versus ionized column densities. Solid black squares and red circles refer to LoBALs and HiBALs, respectively. Open red circles indicate HiBALs with BI$=$0. Bottom panel: BALnicity index versus $\xi$. Solid black squares and red circles refer to LoBALs and HiBALs, respectively. The black stars refer to LoBALs with BI calculated from Mg II. }     
\end{figure}

\section{Results} 
 
To investigate the general X-ray properties of our sample of 39 BALQSOs (9 LoBALs and 30 HiBALs) using the highest quality statistics and compare the results with previous findings, we first measure $N_{\rm{H}}^{n}$  and $\Gamma$ by adopting an absorbed power-law model with neutral absorption. For the remaining 49 sources, upper and lower limits to $N_{\rm{H}}^{n}$ values can be placed using the hardness ratio spectral analysis.  
 
We do not find any correlation between the photon index and intrinsic absorption (Fig.~\ref{fig:gam-nh}).  The mean photon index is $\langle \Gamma\rangle=1.87$, typical of values found in X-ray analyses of radio-quiet type-1 AGN. We detect absorption $N_{\rm{H}}^{n}>5\times$10$^{22}$~cm$^{-2}$ in 25 objects ($\sim$64\%).  If we consider the whole sample of 88 BALQSOs, the fraction of absorbed sources increases to 67\%, still meaning that about one third of the sample is X-ray unabsorbed. We do not find any strong difference between the neutral absorption properties of the ``classical'' (BI$>$0) and those of ``new'' (BI$=$0)  BALQSOs  (Fig.~\ref{fig:nh-numbers}). The Kolmogorov-Smirnov (K--S) test infers the maximum value of the absolute difference between the two cumulative distribution functions, $D=0.25$ with a probability, $p$, that these objects are drawn from the same population of 0.88.

% At least, in the absorption distribution, we can notice that the majority of the sources with BI$=$0 tends to have  $N_{\rm{H}} \sim$10$^{21.5-22.5}$  cm$^{-2}$,
 % except a shallow tendency to measure higher values of   $N_{\rm{H}}$ for ``classical'' BALQSO. 
  
%Few sources present rather flat photon indices $\Gamma$ 1.3 that could indicate the presence of more complex (e.g. ionized/partially covering) absorbers. No significant evolution with redshift is seen. A couple of sources (namely 1007+5343 and 1011+5541)  

Taking into account the difference in spectral and HR analysis strategies (e.g., use of $\chi^2$ instead of \texttt{C-stat}, assuming different energy ranges for the calculation of HR), our results are in very good agreement with \citet{2008A&A...491..425G} for our overlapping 54 BALQSOs.

It is known that the complex structure of a spectrum (presence of ionized/partially covering absorption or additional features in the underlying continuum, such as the soft-excess or scattered component) can conspire to produce apparently lower values of the intrinsic absorption. Taking into account that BALQSOs $are$ sources with complex ionized structures by definition, we fit our 39 objects with the \texttt{absori} model.  
% can be masked by the lower values

This fit of a power law absorbed by ionized material leads to a slight improvement in the `goodness' with respect to the neutral absorbed power-law model, which is however not statistically significant in many cases, mainly because of the poor quality of the data) and provides an acceptable parameterization of the spectra.   It should be emphasized that the need for an ionized absorber (instead of a neutral one) is certainly not required by the X-ray data, but by the likely physical condition of the absorbing gas. forthermore, our purpose is to study and constrain the possible ionized properties of BALQSOs, rather than compare the goodness of fit for different possible models.
%Due to the small numbers of X-ray counts in our sources, we obtained good results even with simple power-law fits to the data.

 In general, the \texttt{absori} model provides higher values of $N_{\rm{H}}^{i}$ for almost all our sources. We detected absorption  $N_{\rm{H}}^{i}>5\times$10$^{21}$~cm$^{-2}$ in 36 sources ($\sim$92\%) and 7 of them showed an absorption $\sim$10$^{23}$~cm$^{-2}$.  A plot of ionized column density versus ionization parameter $\xi$ is given in Fig.~\ref{fig:nh} (top panel).  There is a clear correlation between  $\xi$ and $N_{\rm{H}}^{i}$ for our HiBALs confirmed by both the Kendall's tau ($\tau=0.55$ with a probability of $p=1.99\times$10$^{-5}$) and Spearman's rho ($\rho=0.69$ with $p=2.41\times$10$^{-5}$) tests. 

Some of the LoBALs appear as clear outliers in this graphic and, in general, these sources have lower values of $N_{\rm{H}}^{i}$ than HiBALs. The LoBALs also do not exhibit the stronger $neutral$ absorption that HiBALs do: only in 2 LoBALs (out of 18) do we detect $N_{\rm{H}}^{n}\sim$10$^{23}$ cm$^{-2}$, while most of the objects have lower values ($<$10$^{22}$ cm$^{-2}$). The K--S test confirms that the LoBALs have a different absorption distribution than HiBALs ($p=0.067$ and $p=5.89\times$10$^{-3}$ for the ionized and neutral absorptions, respectively). The  LoBALs-outliers show the highest ionization level of absorbing material (although it is poorly constrained) and the lowest values of absorption among the analysed sources, suggesting that the intrinsic nature of these objects must differ from the remaining BALQSOs. This result agrees with the idea that LoBALs (or at least some of them) may physically differ from HiBALs, and hence require the application of different models for describing their intrinsic properties (e.g.,  partial covering intrinsic absorption).  

While we do not find any strong difference in the neutral absorption properties for our sample of HiBALs as a function of the BALnicity index, we $do$ observe a separation in the case of ionized absorption. For our 6 HiBALs with BI$=$0, we observe mean values of $\langle\xi\rangle=110$ erg cm s$^{-1}$ and $\langle N_{\rm{H}}^{i}\rangle=2.9\times$10$^{22}$ cm$^{-2}$, while the 24 ``classical'' BALQSOs show $\langle\xi\rangle=250$ erg cm s$^{-1}$ and $\langle N_{\rm{H}}^{i}\rangle=8.4\times$10$^{22}$ cm$^{-2}$. The K--S test indicates  only a 10\% chance that these two samples of sources are drawn from the same population.

\begin{figure} 
\centering 
\resizebox{\hsize}{!}{\includegraphics{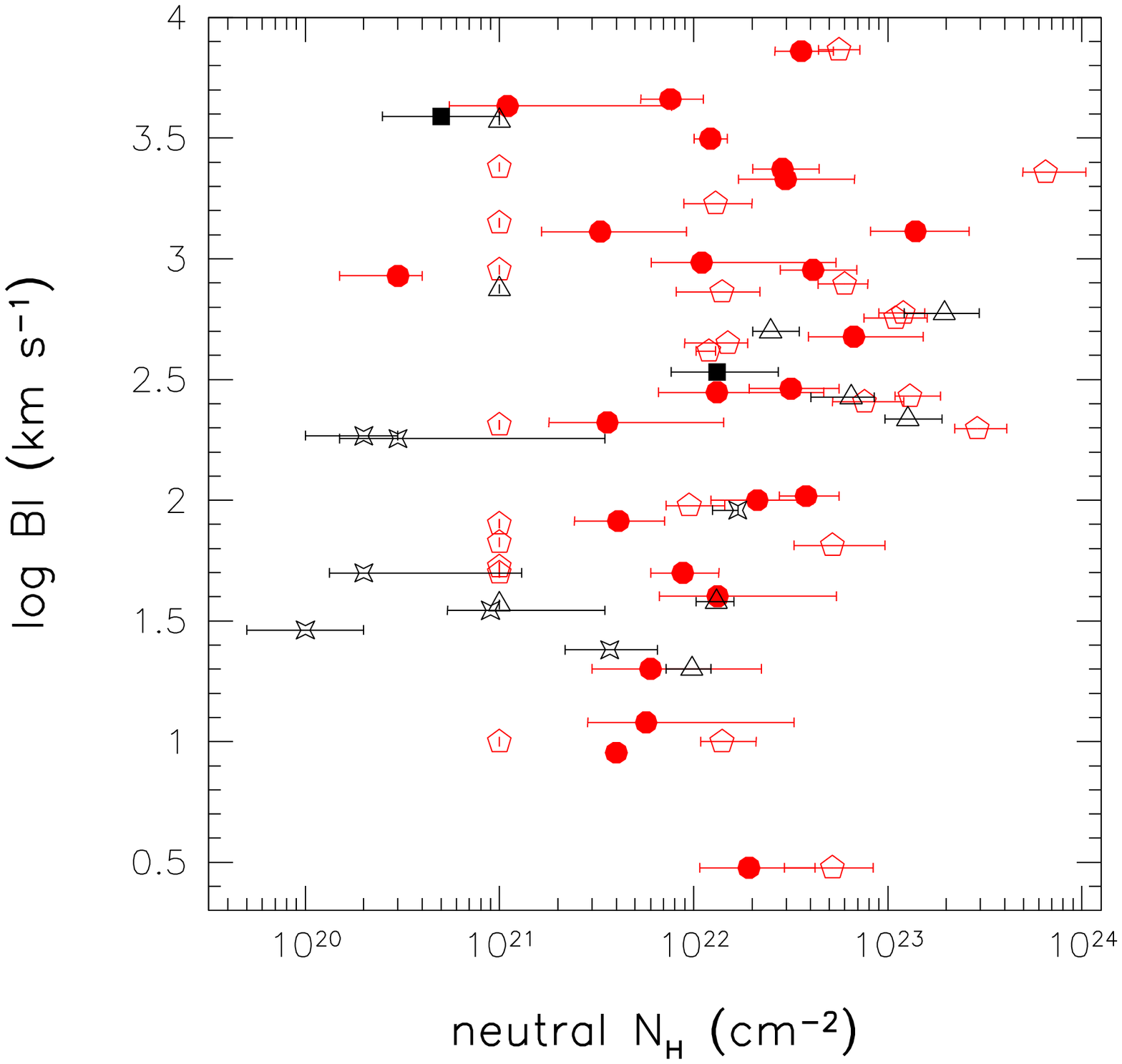}} 
\resizebox{\hsize}{!}{\includegraphics{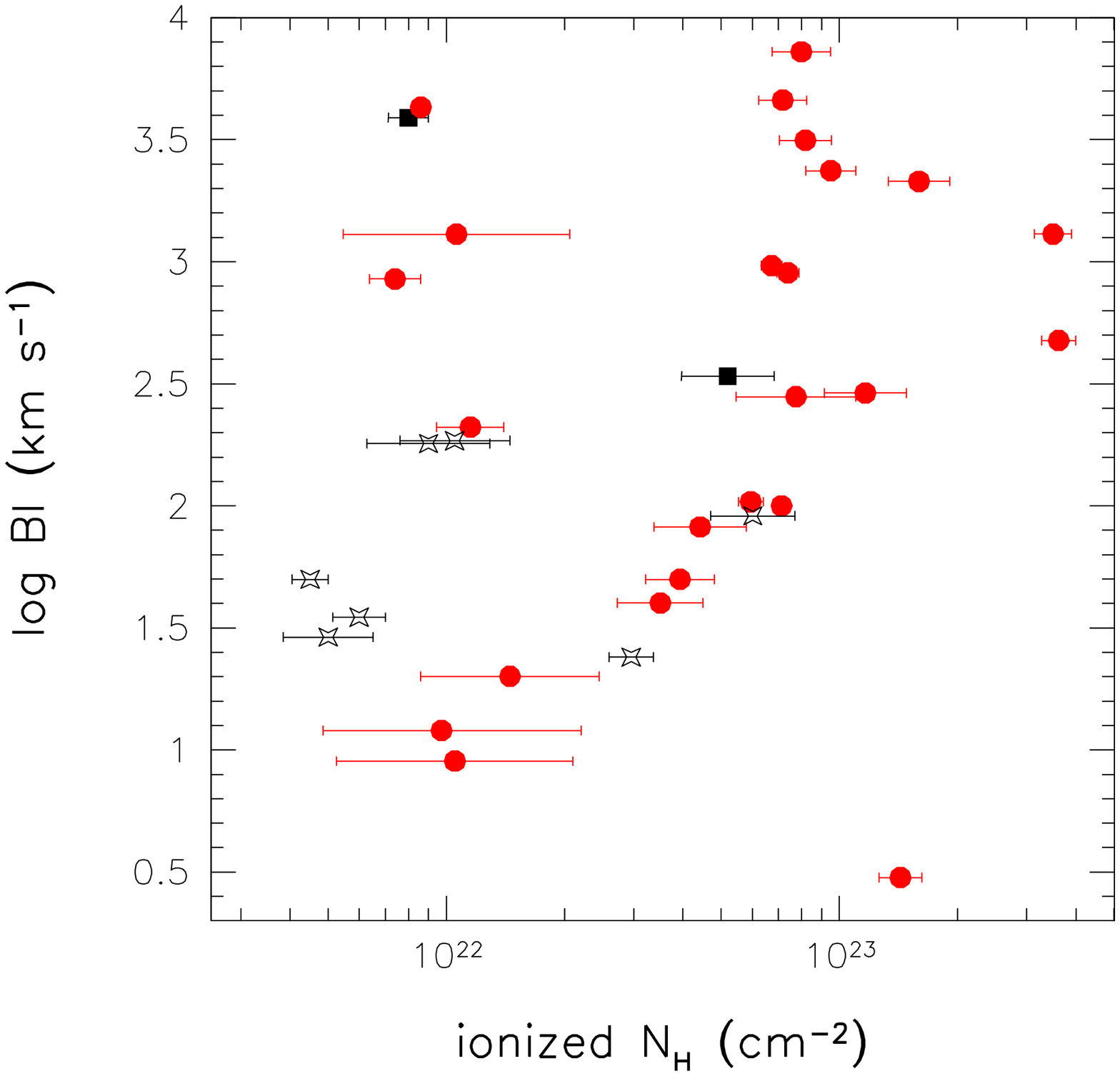}} 
\caption{\label{fig:nh-bi} Dependences of BALnicity index on neutral (top panel) and ionized (bottom panel) absorptions. In both panels, the solid black squares and red circles refer to LoBALs and HiBALs, respectively. The black stars refer to LoBALs with BI calculated from Mg II. Open red pentagons and black triangles refer to HiBALs and LoBALs, whose values of $N_{\rm{H}}$ were obtained from HR analysis.}     
\end{figure}

There is a marginal tendency to measure higher $\xi$ values with increasing BI (Fig.~\ref{fig:nh}, bottom panel). However, both Kendall's tau and Spearman's rho tests give weak support for a correlation. The Kendall's tau method gives $\tau=0.14$ with an associated probability of $p=0.41$, while the Spearman's rank correlation test gives $\rho=0.31$ with $p=0.17$. We also checked the correlations between BAL properties and the neutral/ionized X-ray absorption column density (measured through the X-ray spectral fit or HR analysis). While we do not see any strong dependence between these parameters in the case of neutral absorption (Fig.~\ref{fig:nh-bi}, top panel), there is an apparent trend of increasing ionized $N_{\rm{H}}^{i}$  with BI for HiBALs (Fig.~\ref{fig:nh-bi}, bottom panel).  Applying the Kendall's tau and Spearman's rho tests, we obtain  $\tau=0.32$ and $\rho=0.43$, corresponding to a probability  that the ionized $N_{\rm{H}}^{i}$ and BI are uncorrelated of $p_{\tau}=0.02$  and  $p_{\rho}=0.018$, respectively. Although the significance of the trend is only $\sim$98\%, it might have a very simple physical origin, for which the total amount of outflowing gas largely dictates the BI and the ionized absorption column density.
%that there is no correlation
  
 %correlation between the BI and $\xi$ for HiBALs cannot be rejected

\section{Discussion}

%whether there is strong absorption in these sources or not and that the fraction of the absorbed sources in this class.
As mentioned in the Introduction, an outstanding question about BALQSOs is whether they are all heavily absorbed or only a fraction of them are. Quite controversial results were presented by \citet{2006ApJ...644..709G}  in their Chandra analysis of 35 Large Bright Quasar Survey BALQSOs (all sources have $N_{\rm{H}}^{n} > 10^{22}\:$cm$^{-2}$) and \citet{2008A&A...491..425G}  in their XMM-Newton analysis of 54 SDSS BALQSOs, nearly half of the sources having N$_{\rm{H}}^{n} < 10^{22}\:$cm$^{-2}$. 

Our final sample spans a wide range of intrinsic absorption column densities (derived from the neutral absorption model), although the main result, that $\sim$68\% of 88 BALQSOs have $N_{\rm{H}}^{n}~>~5~\times~10^{21}~\:$cm$^{-2}$, is in agreement with \citet{2008A&A...491..425G}. Thus, the BALQSOs class remains a class of, in general, absorbed sources. For comparison, a typical fraction of broad line AGN with neutral absorption is $\sim$3\% \citep[][]{2009arXiv}, although  BALQSOs seem to be less absorbed than previously assumed.

%Although our samples are partly overlapping, even for sources which are not present in \citet{2008A&A...491..425G} we still observe the same trend - the BALQSOs seem to be less absorbed than previously assumed. 

As mentioned by \citet{2008A&A...491..425G}, some factors can influence this result. One possible explanations may be the different energy range used for the spectral analysis in Chandra and XMM-Newton data (0.5--8 keV and 0.2--10 keV, respectively). In addition,  our work is biased toward the X-ray brightest sources, because we searched for all known BALQSOs with an X-ray detection, while previous works were mainly based on purely optically selected BALQSOs.

Does this result mean that we need to revise our ideas about the nature of BALQSO? The answer is not as obvious as it seems when we use physically more motivated approaches for the modelling of the inner absorption in these objects. Knowing the complicated nature of these sources, we used the \texttt{absori} model to constrain the ionized properties of BALQSOs. Using this model, we detected absorption $N_{\rm{H}}^{i}>5\times$10$^{21}\:$cm$^{-2}$ in $>$90\% of them. The average value of the absorption is in good agreement with the predictions of a radiatively driven accretion disk wind model by \citet{1995ApJ...451..498M}. The observed properties of LoBALs confirm the idea that these objects may physically differ from HiBALs.  

 Using our sample we also have the possibility of comparing properties of ``classical'' (BI$>$0) and ``new'' (BI=0) BALQSOs.  We found that the sources with BI=0 lack significant X-ray absorption, and, in general, this subsample hosts the lowest $N_{\rm{H}}^{i}$ of the entire sample. In our study, the BI$>$0 subsample appears to be more X-ray absorbed and ionized than the BI=0 one, which is clearly consistent with any picture where the mass of outflowing gas largely determines both the optical/UV absorption troughs and the X-ray absorption. In other words, even if we cannot reject the possibility that there are unabsorbed, X-ray bright BALQSOs, it seems that the ``classical'' BALQSOs (i.e., with BI$>$0) are always X-ray absorbed. 

Our results suggest that ``new'' BALQSOs are, likely, a different class of absorbed quasars from the ``classical'' BALQSOs. A similar conclusion is reached by \citet[][]{2008MNRAS.386.1426K}, who studied the  QSO optical spectra in the SDSS DR3 catalogue. The differences in the observational properties of these two classes of objects may be related to the large angle between the line of sight and the axis of the accretion disk plane. In the QSO unification scheme of \citet[][]{2000ApJ...545...63E}, intrinsic absorption lines of different widths form in the same disc wind and are just representations of different lines of sight through the outflowing wind to the quasar continuum source.  The different properties of quasars may also be due to real physical differences between the ``new'' and ``classical'' BALQSOs. In particular, the less ionized and weaker outflows of ``new'' BALQSOs may represent the late evolutionary stages (i.e., dissipation over time) of  ``classical'' massive BAL winds \citep[see review by][]{2004ASPC..311..203H}. In our opinion,  both factors (orientation and evolution of the flow) are likely to contribute to the observed differences between these two classes.

%it may be a result of viewing a common physical structure at different orientation angles.
%Our results demonstrated an obvious dependence between the amount of absorption and the strength of high-ionization UV absorption. the amount of UV absorption and of X-ray absorption $are$ related

A purpose of this paper was also to study the relationship between the BAL properties and X-ray absorption. For the first time, we find a correlation between the ionized absorption and the BALnicity index (i.e., the UV absorption line properties). Although  the correlation is significant at the 98\% level, our results demonstrate that the amount of X-ray absorption and the strength of UV absorption $are$ related when an ionized absorption model is used to model the X-ray data. The physics of this could be as simple as the mass of outflowing gas.

This result may help us to understand the difference between  our  $neutral$ absorbing column density values and those of the \citet{2006ApJ...644..709G}  sample. If we assume a direct dependence between the BALnicity index and absorption, then in the samples with higher BI we must, in general, observe higher X-ray column densities. Although the mean value of the BI in the LBQS BALQSOs sample is $\langle$BI$\rangle=$3437 \kms, in our sample we have just a few objects with BI$>$ 3500 \kms ($\langle$BI$\rangle=$1077 \kms).  The neutral column densities of almost all our high BI objects are $\gtrsim$10$^{22}\:$cm$^{-2}$ and, thus, compatible with the values from \citet{2006ApJ...644..709G}.

Summarizing our results, we can confirm the idea that the presence of the X-ray ionized absorption is important for launching BAL winds. While the influence of other factors on the properties of the wind cannot be rejected, it seems that this dependence is clear: the higher the X-ray absorption, the higher the UV absorption and, hence, the more powerful the outflow. However, the required absorption values for outflow launching mechanisms are not as high as has been previously supposed, which might actually be a selection effect of our X-ray selection bias.

\begin{acknowledgements} 
We thank the anonymous referee for helpful  comments.  The first author would like to thank the Spanish Ministerio de Ciencia e Innovaci\'{o}n for a Juan de la Cierva contract. XB, FJC and RGM acknowledge financial support by the Spanish MICINN under project ESP2006-13608-C02-01.  We also acknowledge the Sloan Digital Sky Survey (\url{http://www.sdss.org/}). Funding for the SDSS has been provided by the Alfred P. Sloan Foundation, the Participating Institutions, the National Science Foundation, the U.S. Department of Energy, the National Aeronautics and Space Administration, the Japanese Monbukagakusho, the Max Planck Society, and the Higher Education Funding Council for England. 
 
\end{acknowledgements} 
 
\bibliographystyle{aa} 
\bibliography{bibbal} 
 
%\Online 
%\begin{appendix} 
%\section{Tables}\label{appendix:tables} 
 
\begin{table*}[!h] 
\caption{2XMM BALQSO sample} 
\label{table:1} 
\centering           
\begin{tabular}{c c c c c c c c} 
\hline\hline 
Name$^{(1)}$ &  2XMM name$^{(2)}$ & $z^{(3)}$  & BAL Type$^{(4)}$ & BI$^{(5)}$ & Analysis$^{(6)}$ &  N. obs.$^{(7)}$& Notes$^{(8)}$\\ 
\hline 
2QZ J003814.3-273130  &  2XMM J003814.1-273131  &  2.405  &  H  &  597  &  HR  &1 & Cr \\ 
SDSS J004206.18-091255.7  &  2XMM J004206.1-091255  &  1.778  &  nHi  &  0  &  HR  &1  & \\ 
SDSS J004338.10+004615.9  &  2XMM J004338.0+004616  &  1.574  &  Lo  &  185  &  A  &1  & Mg II\\ 
SDSS J004341.24+005253.3  &  2XMM J004341.3+005253  &  0.834  &  Lo  &  91  &  A &1 & Mg II\\ 
APMUKS(BJ)B004751.94-523127.9 & 2XMM J005007.3-521506 & 2.422  & Hi & 1300 & A & 3 & C  \\ 
XMMU J010316.7-065137  &  2XMM J010316.2-065135  &  1.914  &  Lo  &  750  &  HR &1  & Br\\ 
XMMU J010328.7-064633  &  2XMM J010328.7-064634  &  1.820  &  Hi  &  210  &  A &1  & Br \\ 
SDSS J010941.97+132843.8  &  2XMM J010941.8+132844  &  1.226  &  Lo  &  20  &  HR  &1  & Mg~II\\ 
$[$WFM91$]$ 0226-1024       &  2XMM J022839.1-101111  &  2.256  &  Hi  &  7344  &  HR  &1  & W\\ 
SDSS J023148.80-073906.3  &  2XMM J023148.8-073905  &  2.512  &  Hi  &  50  &  A &1   & \\ 
SDSS J023224.87-071910.5  &  2XMM J023224.7-071909  &  1.597  &  Hi  &  280  &  A &1  & \\ 
SDSS J024230.65-000029.6  &  2XMM J024230.6-000030  &  2.505  &  H  &  2286  &  HR &1  & \\ 
SDSS J024304.68+000005.4  &  2XMM J024304.6+000005  &  1.995  &  Hi  &  290  &  A &2  & \\ 
SDSS J030222.08+000631.0  &  2XMM J030222.0+000630  &  3.315  &  H  &  100  &  A &1  & \\ 
SDSS J072945.33+370031.9  &  2XMM J072945.3+370032  &  0.969  &  Lo  &  37  &  HR &1  & Mg II\\ 
SDSS J073405.24+320315.2  &  2XMM J073405.2+320315  &  2.082  &  Hi  &  0  &  HR  &1 & \\ 
IRAS J07598+6508          &  2XMM J080430.4+645951  &  0.149  &  Lo  &  3892  &  A  &1 & H\\ 
SDSS J083633.54+553245.0  &  2XMM J083633.4+553245  &  1.614  &  Hi  &  0  &  HR  &1 &  \\ 
CXOSEXSI J084840.5+445732  &  2XMM J084840.5+445733  &  3.093  &  Hi &  2136  &  A  &3 & St\\ 
SDSS J085551.24+375752.2  &  2XMM J085551.1+375752  &  1.930  &  Hi  &  1296  &  A &1 & \\ 
SDSS J091127.61+055054.1  &  2XMM J091127.5+055054  &  2.763  &  H  &  2358  &  A  &1 & \\ 
SDSS J091914.23+303019.0  &  2XMM J091914.2+303018  &  1.387  &  Lo  &  38  &  HR  &1 & Mg II\\ 
SDSS J092104.36+302030.3  &  2XMM J092104.3+302031  &  3.350  &  H  &  40  &  A  &3 & \\ 
SDSS J092138.45+301546.9  &  2XMM J092138.4+301546  &  1.589  &  Hi  &  0  &  HR &1 & \\ 
SDSS J092142.57+513149.4  &  2XMM J092142.5+513148  &  1.835  &  nHi  &  67  &  HR&1  & \\ 
SDSS J092238.43+512121.2  &  2XMM J092238.3+512120  &  1.751  &  Lo  &  180  &  A &1 & Mg II\\ 
SDSS J092345.19+512710.0  &  2XMM J092345.1+512711  &  2.164  &  Hi  &  1696  &  HR&1  & \\ 
SDSS J093918.07+355615.0  &  2XMM J093918.1+355612  &  2.047  &  Hi  &  80  &  HR &1 & \\ 
SDSS J094309.56+481140.5  &  2XMM J094309.8+481142  &  1.809  &  Hi  &  65  &  HR &1 & \\ 
SDSS J095548.13+410955.3  &  2XMM J095548.1+410955  &  2.308  &  nH  &  10  &  HR &1 & \\ 
SDSS J095942.08+024103.1  &  2XMM J095942.0+024104  &  1.795  &  Lo  &  595  &  HR &1 & P\\ 
SDSS J095944.47+051158.3  &  2XMM J095944.4+051157  &  1.596  &  Hi  &  0  &  HR &1 & \\ 
SDSS J100038.01+020822.3  &  2XMM J100037.9+020822  &  1.825  &  Lo  &  267  &  HR&1  & P\\ 
SDSS J100116.78+014053.5  &  2XMM J100116.7+014053  &  2.055  &  Hi  &  0  &  A  &2& \\ 
SDSS J100120.84+555349.5  &  2XMM J100120.7+555351  &  1.413  &  Lo  &  27  &  A &2 & Mg II\\ 
SDSS J100145.15+022456.9  &  2XMM J100145.2+022456  &  2.032  &  nHi  &  10  &  HR &2 & \\ 
PG 1004+130              &  2XMM J00726.0+124856  &  0.241  &  Hi  &  850  &  A &1 & Wi\\ 
SDSS J100728.69+534326.7  &  2XMM J100728.8+534327  &  1.768  &  nHi  &  3  &  A &1 & \\ 
SDSS J101144.33+554103.1  &  2XMM J101144.4+554103  &  2.811  &  H  &  0  &  A &1 & \\ 
SDSS J101614.26+520915.7  &  2XMM J101614.2+520916  &  2.455  &  H  &  2401  &  HR &1& Gr\\ 
SDSS J101954.54+082515.0  &  2XMM J101954.6+082515  &  3.010  &  nH  &  0  &  HR &1& \\ 
SDSS 102117.74+131545.9  &  2XMM J102117.7+131546  &  1.565  &  Lo  &  35  &  A &1& Mg II\\ 
SDSS J104433.04-012502.2  &  2XMM J104433.0-012500  &  5.800  &  Hi  &  900  &  HR &1& G\\ 
ISO${\_}$LHDS J105154+572408 &  2XMM J105154.5+572407  &  2.400  &  Hi  &  900  &  A & 12 & LH\\ 
SDSS J105201.35+441419.8  &  2XMM J105201.3+441417  &  1.791  &  Hi  &  476  &  A &1& \\ 
RX J105207.7+573842      &  2XMM J105207.4+573838  &  2.730  &  Hi  &  4300  &  A&12 & LH\\ 
SDSS J110637.16+522233.4  &  2XMM J110637.0+522233  &  1.840  &  Hi  &  0  &  HR  &1& \\ 
SDSS J110853.98+522337.9  &  2XMM J110853.5+522341  &  1.664  &  Hi  &  50  &  HR &1 & \\ 
$[$HB89$]$ 1115+080          &  2XMM J111816.9+074558  &  1.735  &  Hi  &  9  &  A  &3& \\ 
SDSS J111859.56+075606.5  &  2XMM J111859.7+075602  &  1.760  &  Hi  &  3  &  HR  &1& \\ 
SDSS J112020.96+432545.1  &  2XMM J112020.9+432545  &  3.548  &  H  &  0  &  A  &1& \\ 
SDSS J112300.25+052451.0  &  2XMM J112300.3+052448  &  3.700  &  nH  &  0  &  HR &1 & \\ 
UM 425                   &  2XMM J112320.7+013748  &  1.462  &  Hi  &  415  &  HR &1 & Hu \\ 
SDSS J112432.14+385104.3  &  2XMM J112432.0+385104  &  3.530  &  H  &  270  &  HR &1 & \\ 
SDSS J113537.67+491323.2  &  2XMM J113537.6+491322  &  1.982  &  nH  &  205  &  HR &1 & \\ 
SDSS J114111.61-014306.6  &  2XMM J114111.5-014305  &  1.266  &  Lo  &  500  &  HR &1 & Mg II\\ 
SDSS J114636.88+472313.3  &  2XMM J114636.9+472313  &  1.895  &  Lo  &  24  &  A &1 & \\ 
SDSS J120522.18+443140.4  &  2XMM J120522.1+443141  &  1.921  &  Hi  &  965  &  A &1 & \\ 
SDSS J122307.52+103448.2  &  2XMM J122307.4+103448  &  2.742  &  H  &  0  &  A  &1& \\ 
SDSS J122637.02+013015.9  &  2XMM J122636.9+013016  &  1.552  &  Hi  &  787  &  HR &1 & \\ 
SDSS J122708.29+012638.4  &  2XMM J122708.2+012638  &  1.954  &  Hi  &  730  &  HR &1 & \\ 
SDSS J123637.45+615814.4  &  2XMM J123637.7+615813  &  2.520  &  H  &  447  &  HR &1 & \\ 
LBQS 1235+1807B          &  2XMM J123820.3+175038  &  0.449  &  Lo &  5086  &  HR &1 & Mg II,  F\\ 
SDSS J124520.72-002128.2  &  2XMM J124520.6-002127  &  2.354  &  H  &  0  &  A &1 & \\ 
SDSS J124559.59+570053.1  &  2XMM J124559.6+570052  &  1.656  &  Hi  &  0  &  HR &1 & \\ 
\hline 
\end{tabular}\\ 
\end{table*}

\addtocounter{table}{-1} 
\begin{table*}[!h] 
\caption{2XMM BALQSO sample} 
\centering           
\begin{tabular}{c c c c c c c c} 
\hline\hline 
Name$^{(1)}$ &  2XMM name$^{(2)}$ & $z^{(3)}$  & BAL Type$^{(4)}$ & BI$^{(5)}$ & Analysis$^{(6)}$ &  N. obs.$^{(7)}$& Notes$^{(8)}$\\ 
 
\hline 
 
$[$HB89$]$ 1246-057          &  2XMM J124913.8-055918  &  2.236  &  Hi  &  4590  &  A &2 & J\\ 
SDSS J132330.44+545955.6  &  2XMM J132330.7+545954  &  2.208  &  Hi  &  0  &  A &1 & \\ 
SDSS J132401.53+032020.6  &  2XMM J132401.4+032019  &  0.926  &  nLo  &  217  &  HR &1 & Mg II\\ 
SDSS J132827.07+581836.9  &  2XMM J132827.3+581839  &  3.139  &  H  &  198  &  HR &1 & \\ 
SDSS J133553.61+514744.1  &  2XMM J133553.7+514744  &  1.830  &  Lo  &  340  &  A  &1& \\ 
SDSS J133639.40+514605.2  &  2XMM J133639.1+514608  &  2.229  &  Hi  &  1410  &  HR &1 & \\ 
SDSS J134059.24-001944.9  &  2XMM J134059.1-001945  &  1.857  &  nHi  &  0  &  HR  &1& \\ 
HELLAS2XMM J140049.1+025850  &  2XMM J140048.7+025852  &  1.822  &  Hi  &  569  &  HR &2 & C\\ 
SDSS J142539.38+375736.7  &  2XMM J142539.3+375736  &  1.897  &  Hi  &  104  &  A  &1& \\ 
SDSS J142555.22+373900.7  &  2XMM J142555.2+373900  &  2.691  &  H  &  53  &  HR &1 & \\ 
SDSS J142652.94+375359.9  &  2XMM J142652.8+375401  &  1.812  &  Hi  &  95  &  HR &1 & \\ 
SDSS J143513.89+484149.3  &  2XMM J143513.9+484149  &  1.886  &  Hi  &  20  &  A &2 & \\ 
SDSS J144625.48+025548.6  &  2XMM J144625.6+025549  &  1.866  &  Hi  &  256  &  HR &1 & \\ 
SDSS J144727.49+403206.3  &  2XMM J144727.4+403206  &  1.335  &  Lo  &  50  &  A  &1& Mg II\\ 
SDSS J150858.15+565226.5  &  2XMM J150858.2+565227  &  1.797  &  Hi  &  0  &  HR  &1& \\ 
SDSS J151729.70+001652.6  &  2XMM J151729.6+001651  &  1.887  &  nHi  &  80  &  HR &1 & \\ 
SDSS J152553.89+513649.1  &  2XMM J152553.8+513649  &  2.883  &  Hi  &  3144  &  A &3 & \\ 
SDSS J153322.80+324351.4  &  2XMM J153322.8+324351  &  1.899  &  Hi  &  0  &  HR  &1& \\ 
SDSS J154359.44+535903.2  &  2XMM J154359.4+535902  &  2.370  &  Hi  &  82  &  A  &2& \\ 
SDSS J203941.04-010201.6  &  2XMM J203941.2-010202  &  2.065  &  Hi  &  0  &  HR  &1& \\ 
LBQS 2111-4335    &    2XMM J211507.0-432310  & 1.708   &  Hi  &   7249   &  A  &1& Mor\\  
SDSS J213023.61+122252.2  &  2XMM J213023.4+122251  &  3.263  &  nH  &  12  &  A  &1& \\ 
SDSS J231850.79+002552.6  &  2XMM J231850.6+002554  &  1.591  &  Hi  &  0  &  HR  &1& \\ 
 
\hline 
\end{tabular}\\ 
\begin{flushleft} 
\small{{\bf Notes}: Col.(1): Source name; Col. (2): 2XMM source name; Col.(3): Redshift, taken from SDSS or NED; Col. (4): The BALQSO subclassification: Hi - HiBAL,  Lo - LoBAL, H - HiBAL in which the MgII region is not within the spectral coverage (i.e., impossible to check the presence of low-ionization absorption troughs), n  - relatively narrow trough; Col.(5): BI - BALnicity Index (in units of \kms); Col.(6): X-ray analysis used in this work: A $-$ spectral analysis, HR $-$ hardness ratio analysis; Col.(7): Number of individual XMM-Newton observations; Col.(8): Notes on individual objects: ``Mg~II'' denotes LoBAL, which BALnicity index were computed from Mg~II absorption troughs; ``Cr''  - the optical spectrum was retrieved from \citet{2004MNRAS.349.1397C}; ``C'' -  optical spectra courtesly provided by \citet{2007A&A...466...31C}; ``Br'' -  optical spectra from AXIS survey courtesly provided by \citet{2002A&A...382..522B}; ``W'' -  BALnicity index value was taken from \citet{1991ApJ...373...23W};  ``H'' - reconstructed spectrum  from \citet{1995ApJ...448L..69H}; ``St'' - reconstructed spectrum from \citet{2002AJ....123.2223S}; ``P'' -  optical spectra courtesly provided by \citet{2006ApJ...644..100P}; ``Wi'' - BALnicity index value was taken from \citet{1999ApJ...520L..91W}; ``Gr'' - BALnicity index value was taken from \citet{2000ApJ...544..142G}; ``G''  - BALnicity index value was taken from \citet{2001ApJ...561L..23G}; ``LH'' - spectra were taken and reconstructed from \citet{2001A&A...371..833L}; ``Hu'' - BALnicity index value was taken from \citet{2000A&A...358..835H}; ``F'' - reconstructed spectrum from \citet{1987AJ.....94.1423F}; ``J'' - reconstructed spectrum from \citet{1987ApJ...317..460J}; ``Mor'' - reconstructed spectrum from \citet{1991AJ....102.1627M}. The spectra for rest of the objects were retrieved from the SDSS archive and correspond to the following catalogues: \citet{2006ApJS..165....1T}; \citet{2008ApJ...680..169S} and \citet{2009ApJ...692..758G}.} 
\end{flushleft} 
\end{table*}

\begin{table*} 
\caption{Best fit parameters from the absorbed power-law model for a sample of 39 BALQSOs with the best statistics.} 
\label{table:2} 
\centering           
\begin{tabular}{ c c c c c c c c} 
\hline\hline 
Name$^{(1)}$ & $N_{\rm{H, Gal}}$$^{(2)}$ & $\Gamma$$^{(3)}$ & $N_{\rm{H}}$$^{(4)}$  & $\log f_X(0.5-2)$$^{(5)}$ & $\log f_X(2-10)$$^{(6)}$ & $\log L(0.5-2)$$^{(7)}$& $\log L(2-10)$$^{(8)}$\\ 
 
\hline  
 
J004338.10+004615.9  & 1.80 & 1.77 $^{+ 0.14 }_{ -0.13 }$ & 0.02 $^{+ 0.01 }_{ -0.02 }$ & -13.59 & -13.35 & 44.42 & 44.77 \\ 
J004341.24+005253.3  & 1.80 & 1.89 $^{+ 0.33 }_{ -0.30 }$ & 1.69 $^{+ 0.73 }_{ -0.59 }$ & -13.71 & -13.19 & 44.11 & 44.33 \\ 
B004751.94-523127.9  & 2.44 & 2.32 $^{+ 0.96 }_{ -0.82 }$ & 13.90 $^{+ 12.29 }_{ -9.70 }$ & -14.42 & -14.02 & 44.64 & 44.86 \\ 
J010328.7-064633 &  5.70 & 2.17 $^{+ 0.66 }_{ -0.52 }$ & 0.36 $^{+ 1.07 }_{ -0.36 }$ & -13.48 & -13.41 & 44.62 & 44.99 \\ 
J023148.80-073906.3 & 3.32 & 2.08 $^{+ 0.35 }_{ -0.30 }$ & 0.88 $^{+ 0.47 }_{ -0.41 }$ & -13.82 & -13.71 & 44.60 & 45.03 \\ 
J023224.87-071910.5 & 3.02 & 1.65 $^{+ 1.00 }_{ -0.67 }$ & 1.32 $^{+ 3.35 }_{ -1.32 }$ & -14.04 & -13.57 & 44.01 & 44.52 \\ 
J024304.68+000005.4 & 3.07 & 1.73 $^{+ 0.19 }_{ -0.18 }$ & 3.17 $^{+ 2.43 }_{ -2.02 }$ & -13.79 & -13.30 & 44.52 & 45.04 \\ 
J030222.08+000631.0& 6.79 & 1.71 $^{+ 0.27 }_{ -0.25 }$ & 2.13 $^{+ 2.00 }_{ -1.56 }$ & -14.11 & -13.72 & 44.01 & 45.04 \\ 
07598+6508 & 4.18 & 2.72 $^{+ 0.32 }_{ -0.17 }$ & 0.05 $^{+ 0.05 }_{ -0.05 }$ & -13.63 & -13.95 & 42.18 & 41.87 \\ 
J084840.5+445732 & 2.79 & 2.03 $^{+ 0.55 }_{ -0.43 }$ & 2.98 $^{+ 3.77 }_{ -2.23 }$ & -14.24 & -14.04 & 44.38 & 44.89 \\ 
J085551.24+375752.2 & 3.15 & 1.60 $^{+ 0.31 }_{ -0.26 }$ & 0.33 $^{+ 0.59 }_{ -0.33 }$ & -13.89 & -13.51 & 44.14 & 44.73 \\ 
J091127.61+055054.1 &  3.36 & 1.62 $^{+ 0.17 }_{ -0.16 }$ & 2.87 $^{+ 1.56 }_{ -1.22 }$ & -13.29 & -12.80 & 44.95 & 45.73 \\ 
J092104.36+302030.3  & 1.67 & 2.02 $^{+ 0.84 }_{ -0.50 }$ & 1.33 $^{+ 4.11 }_{ -1.32 }$ & -14.16 & -14.04 & 44.54 & 44.98 \\ 
J092238.43+512121.2 &  1.32 & 2.18 $^{+ 0.33 }_{ -0.20 }$ & 0.03 $^{+ 0.32 }_{ -0.03 }$ & -13.74 & -13.76 & 44.56 & 44.65 \\ 
J100116.78+014053.5 &  1.94 & 1.90 $^{+ 0.13 }_{ -0.11 }$ & 0.04 $^{+ 0.14 }_{ -0.04 }$ & -13.92 & -13.77 & 44.33 & 44.68 \\ 
J100120.84+555349.5 &  0.89 & 2.11 $^{+ -0.04 }_{ -0.04 }$ & 0.01 $^{+ 0.01 }_{ -0.01 }$ & -12.07 & -13.06 & 46.02 & 46.08 \\ 
1004+130 &  3.54 & 1.40 $^{+ 0.04 }_{ -0.04 }$ & 0.03 $^{+ 0.01 }_{ -0.03 }$ & -12.99 & -12.48 & 43.19 & 43.71 \\ 
J100728.69+534326.7 &  0.71 & 1.85 $^{+ 0.73 }_{ -0.58 }$ & 1.93 $^{+ 2.30 }_{ -1.53 }$ & -14.02 & -13.67 & 44.38 & 44.62 \\ 
J101144.33+554103.1 &  0.76 & 1.18 $^{+ 0.38 }_{ -0.34 }$ & 1.60 $^{+ 3.06 }_{ -1.08 }$ & -14.29 & -13.64 & 43.99 & 44.72 \\ 
J102117.74+131545.9 &  4.01 & 2.07 $^{+ 0.22 }_{ -0.19 }$ & 0.09 $^{+ 0.26 }_{ -0.06 }$ & -13.36 & -13.27 & 44.58 & 44.95 \\ 
J105154+572408 & 0.56 & 2.18 $^{+ 0.52 }_{ -0.40 }$ & 4.13 $^{+ 2.80 }_{ -1.96 }$ & -13.63 & -13.43 & 44.28 & 44.35 \\ 
J105201.35+441419.8 & 1.04 & 1.81 $^{+ 1.00 }_{ -0.81 }$ & 6.70 $^{+ 8.56 }_{ -4.79 }$ & -14.05 & -13.44 & 44.56 & 44.85 \\ 
J105207.7+573842 &  0.56 & 1.74 $^{+ 0.25 }_{ -0.17 }$ & 0.11 $^{+ 0.66 }_{ -0.11 }$ & -14.51 & -14.27 & 44.05 & 44.39 \\ 
1115+080 &  3.57 & 1.87 $^{+ 0.02 }_{ -0.02 }$ & 0.40 $^{+ 0.03 }_{ -0.03 }$ & -12.67 & -12.43 & 45.39 & 45.83 \\ 
J112020.96+432545.1 & 2.90 & 1.56 $^{+ 0.33 }_{ -0.28 }$ & 0.95 $^{+ 1.97 }_{ -0.95 }$ & -13.95 & -13.54 & 44.37 & 45.22 \\ 
J114636.88+472313.3 &  2.41 & 2.02 $^{+ 0.19 }_{ -0.18 }$ & 0.37 $^{+ 0.28 }_{ -0.26 }$ & -13.17 & -13.05 & 44.08 & 44.37 \\ 
J120522.18+443140.4 &  1.15 & 1.42 $^{+ 0.90 }_{ -0.35 }$ & 1.10 $^{+ 4.30 }_{ -0.90 }$ & -14.17 & -13.59 & 43.93 & 44.54 \\ 
J122307.52+103448.2 & 2.21 & 2.18 $^{+ 0.19 }_{ -0.18 }$ & 0.78 $^{+ 0.55 }_{ -0.51 }$ & -13.39 & -13.37 & 45.21 & 45.53 \\ 
J124520.72-002128.2 &  1.69 & 1.86 $^{+ 0.45 }_{ -0.36 }$ & 4.04 $^{+ 3.71 }_{ -2.13 }$ & -14.18 & -13.80 & 44.39 & 44.78 \\ 
1246-057 &  2.07 & 1.61 $^{+ 0.13 }_{ -0.12 }$ & 0.76 $^{+ 0.36 }_{ -0.32 }$ & -13.85 & -13.47 & 44.36 & 44.92 \\ 
J132330.44+545955.6 &  1.60 & 1.61 $^{+ 0.29 }_{ 0.27 }$ & 1.10 $^{+ 1.14 }_{ -0.98 }$ & -13.36 & -12.95 & 44.92 & 45.43 \\ 
J133553.61+514744.1 &  0.96 & 2.04 $^{+ 0.53 }_{ -0.43 }$ & 1.32 $^{+ 1.41 }_{ -0.95 }$ & -14.11 & -13.92 & 44.34 & 44.49 \\ 
J142539.38+375736.7 & 1.04 & 2.53 $^{+ 0.55 }_{ -0.45 }$ & 3.80 $^{+ 1.80 }_{ -1.42 }$ & -13.70 & -13.62 & 45.18 & 45.06 \\ 
J143513.89+484149.3 & 2.96 & 1.58 $^{+ 0.43 }_{ -0.37 }$ & 0.60 $^{+ 1.64 }_{ -0.60 }$ & -13.63 & -13.19 & 44.43 & 45.02 \\ 
J144727.49+403206.3 &  1.17 & 2.16 $^{+ 0.20 }_{ -0.14 }$ & 0.02 $^{+ 0.11 }_{ -0.01 }$ & -13.44 & -13.46 & 44.59 & 44.64 \\ 
J152553.89+513649.1 & 1.62 & 1.83 $^{+ 0.07 }_{ -0.07 }$ & 1.22 $^{+ 0.27 }_{ -0.26 }$ & -13.06 & -12.81 & 45.48 & 45.94 \\ 
J154359.44+535903.2 &  1.23 & 1.71 $^{+ 0.12 }_{ -0.12 }$ & 0.41 $^{+ 0.30 }_{ -0.28 }$ & -13.29 & -13.00 & 45.07 & 45.50 \\ 
2111-4335 &   3.30 &  1.81 $^{+ 0.29 }_{ -0.27 }$ & 3.58 $^{+ 1.68 }_{ -1.30 }$ & -13.60 & -13.09 & 44.69 & 45.14 \\ 
J213023.61+122252.2 & 6.55 & 1.98 $^{+ 0.50 }_{ -0.37 }$ & 0.57 $^{+ 2.72 }_{ -0.57 }$ & -13.57 & -13.38 & 44.66 & 44.52 \\

\hline 
\end{tabular}\\ 
\begin{flushleft} 
\small{Notes: Col.(1): Source name; Col.(2): Galactic neutral hydrogen column density in units of $10^{20}$ cm$^{-2}$ , taken from \citet{1990ARA&A..28..215D}; Col.(3): Photon Index; Col.(4): Neutral hydrogen column density, in units of $10^{22}$ cm$^{-2}$; Col. (5) and Col. (6): Logarithms of $0.5-2$ and $2-10$~keV observed fluxes, in units of erg s$^{-1}$ cm$^{-2}$; Col.(7) and Col.(8): Logarithms of $0.5-2$ and $2-10$~keV rest-frame luminosities, corrected for intrinsic absorption, in units of erg s$^{-1}$.} 
\end{flushleft} 
\end{table*} 
%\clearpage 
 
\begin{table*} 
\caption{\texttt{absori} model fits to the X-ray spectra for a sample of 39 BALQSOs with the best statistics.} 
\label{table:3} 
\centering           
\begin{tabular}{ c c c c c} 
\hline\hline 
Name$^{(1)}$ & $\Gamma$$^{(2)}$ & BAL type$^{(3)}$& $\xi$$^{(4)}$ & $N_{\rm{H}}$$^{(5)}$  \\ 
\hline  
 
J004338.10+004615.9& 1.77 & Lo & 450 $^{+300}_{-300 }$ & 1.05$^{+0.40}_{-0.40}$ \\ 
J004341.24+005253.3 & 1.89 & Lo & 150 $^{+100}_{-100 }$ & 6.02$^{+1.69}_{-1.69}$ \\ 
B004751.94-523127.9 & 2.32 & Hi & 450 $^{+250}_{-250 }$ & 34.97$^{+4.00}_{-4.00}$ \\ 
J010328.7-064633 & 2.17 & Hi & 150 $^{+120}_{-120 }$ & 1.15$^{+0.25}_{-0.25}$ \\ 
J023148.80-073906.3 & 2.08 & Hi & 300 $^{+200}_{-200 }$ & 3.93$^{+0.88}_{-0.88}$ \\ 
J023224.87-071910.5 & 1.65 & Hi & 300 $^{+200}_{-200 }$ & 7.75$^{+3.25}_{-3.25}$ \\ 
J024304.68+000005.4 & 1.73 & Hi & 205 $^{+195}_{-195 }$ & 11.65$^{+3.16}_{-3.16}$ \\ 
J030222.08+000631.0 & 1.71 & H & 100 $^{+25}_{-25 }$ & 7.12$^{+0.35}_{-0.35}$ \\ 
J07598+6508 & 2.72 & Lo & 500 $^{+300}_{-300}$ & 0.80$^{+0.10}_{-0.10}$ \\ 
J084840.5+445732 & 2.03 & Hi & 500 $^{+300}_{-300 }$ & 15.94$^{+3.15}_{-3.15}$ \\ 
J085551.24+375752.2 & 1.60 & Hi & 100 $^{+50}_{-50 }$ & 1.06$^{+1.00}_{-1.00}$ \\ 
J091127.61+055054.1 & 1.62 & H & 275 $^{+125}_{-125 }$ & 9.50$^{+1.50}_{-1.50}$ \\ 
J092104.36+302030.3 & 2.02 & H & 100 $^{+100}_{-100 }$ & 3.5$^{+1.0}_{-1.0}$ \\ 
J092238.43+512121.2 & 2.18 & Lo & 500 $^{+200}_{-200 }$ & 0.90$^{+0.39}_{-0.39}$ \\ 
J100116.78+014053.5 & 1.90 & Hi & 50 & 0.17$^{+0.05}_{-0.05}$ \\ 
J100120.84+555349.5 & 2.11 & Lo & 600 $^{+100}_{-100 }$ & 0.50$^{+0.15}_{-0.15}$ \\ 
 1004+130           & 1.40 & Hi & 225 $^{+75}_{-75 }$ & 0.74$^{+0.12}_{-0.12}$ \\ 
J100728.69+534326.7 & 1.85 & nHi & 425 $^{+125}_{-125 }$ & 14.31$^{+1.92}_{-1.92}$ \\ 
J101144.33+554103.1 & 1.18 & H & 100 $^{+50}_{-50 }$ & 2.10$^{+0.40}_{-0.40}$ \\ 
J102117.74+131545.9 & 2.07 & Lo & 75 $^{+25}_{-25 }$ & 0.60$^{+0.10}_{-0.10}$ \\ 
J105154+572408 & 2.18 & Hi & 50$^{+10}_{-10}$ & 7.40$^{+0.50}_{-0.50}$ \\ 
J105201.35+441419.8 & 1.81 & Hi & 650 $^{+150}_{-150 }$ & 36.15$^{+3.75}_{-3.75}$ \\ 
J105207.7+573842 & 1.74 & Hi & 50 & 0.86 \\ 
 1115+080      & 1.87 & Hi & 10 $^{+7.7}_{-13 }$ & 1.05$^{+1.05}_{-1.05}$ \\ 
J112020.96+432545.1 & 1.56 & H & 75 $^{+25}_{-25 }$ & 0.98$^{+0.18}_{-0.18}$ \\ 
J114636.88+472313.3 & 2.02 & Lo & 110 $^{+90}_{-90 }$ & 2.95$^{+0.41}_{-0.41}$ \\ 
J120522.18+443140.4 & 1.42 & Hi & 350 $^{+50}_{-50 }$ & 6.72$^{+0.42}_{-0.42}$ \\ 
J122307.52+103448.2 & 2.18 & H & 200 $^{+50}_{-50 }$ & 2.65$^{+0.25}_{-0.25}$ \\ 
J124520.72-002128.2 & 1.86 & H & 100& 7.90\\ 
1246-057 & 1.61 & Hi & 400 $^{+200}_{-200 }$ & 7.17$^{+1.08}_{-1.08}$ \\ 
J132330.44+545955.6 & 1.61 & Hi & 150 $^{+150}_{-150 }$ & 3.65$^{+1.35}_{-1.35}$ \\ 
J133553.61+514744.1 & 2.04 & Lo & 250 $^{+150}_{-150 }$ & 5.20$^{+1.62}_{-1.62}$ \\ 
J142539.38+375736.7 & 2.53 & Hi & 100 $^{+50}_{-50 }$ & 5.95$^{+0.45}_{-0.45}$ \\ 
J143513.89+484149.3 & 1.58 & Hi & 100 $^{+50}_{-50 }$ & 1.45$^{+1.00}_{-1.00}$ \\ 
J144727.49+403206.3 & 2.16 & Lo & 125 $^{+75}_{-75 }$ & 0.45$^{+0.05}_{-0.05}$ \\ 
J152553.89+513649.1 & 1.83 & Hi & 375 $^{+175}_{-175 }$ & 8.20$^{+1.35}_{-1.35}$ \\ 
J154359.44+535903.2 & 1.71 & Hi & 300 $^{+300}_{-300 }$ & 4.42$^{+1.37}_{-1.37}$ \\ 
2111-4335 & 1.81 &  Hi & 100 $^{+50}_{-50 }$ & 8.00$^{+1.52}_{-1.52}$ \\ 
J213023.61+122252.2 & 1.98 & nH & 10 & 0.97$^{+0.97}_{-1.23}$ \\ 
\hline 
\end{tabular}\\ 
\begin{flushleft} 
\small{Notes: Col.(1): Source name; Col.(2): Photon Index (fixed), taken from the absorbed power-law model fits; Col.(3): The BALQSO subclassification; Col.(4): Ionisation parameter in erg cm s$^{-1}$; Col.(5): Hydrogen column density of absorber in units of $10^{22}$ cm$^{-2}$.} 
\end{flushleft} 
\end{table*}

\begin{table*} 
\caption{\label{table:4}Hardness ratio analysis results} 
\centering           
\begin{tabular}{ c c c c } 
\hline\hline 
Name$^{(1)}$  & $N_{\rm{H}}$$^{(2)}$  \\ 
\hline  
2QZ J003814.3-273130 & 12.05$^{+3.50 }_{-4.10 }$ \\ 
SDSS J004206.18-091255.7 & 3.00$^{+2.2 }_{ -2.00 }$ \\ 
XMMU J010316.7-065137 & $<0.10 $\\ 
SDSS J010941.97+132843.8 & 0.98$^{+0.25 }_{ -0.35 }$ \\ 
$[$WFM91$]$ 0226-1024 & 5.60$^{+1.60 }_{ -1.50 }$ \\ 
SDSS J024230.65-000029.6 & 65.00$^{+40.00 }_{ -20.00 }$ \\ 
SDSS J072945.33+370031.9 & $<0.10 $\\ 
SDSS J073405.24+320315.2 & 1.30$^{+0.20 }_{ -0.4 }$ \\ 
SDSS J083633.54+553245.0 & 1.55$^{+0.25 }_{ -0.35 }$ \\ 
SDSS J091914.23+303019.0 & 1.30$^{+0.30 }_{ -0.35 }$ \\ 
SDSS J092138.45+301546.9 & 8.80$^{+3.40 }_{ -3.60 }$ \\ 
SDSS J092142.57+513149.4 & $<0.10 $\\ 
SDSS J092345.19+512710.0 & 1.30$^{+0.70 }_{ -0.60 }$ \\ 
SDSS J093918.07+355615.0 & $<0.10 $\\ 
SDSS J094309.56+481140.5 & 5.20$^{+4.50 }_{-3.00 }$ \\ 
SDSS J095548.13+410955.3 & $<0.10 $\\ 
SDSS J095942.08+024103.1 & 19.60$^{+10.00}_{ -12.00 }$ \\ 
SDSS J095944.47+051158.3 & 6.90$^{+2.50 }_{ -3.00 }$ \\ 
SDSS J100038.01+020822.3 & 6.50$^{+2.00 }_{ -4.00 }$ \\ 
SDSS J100145.15+022456.9 & 1.40$^{+0.70 }_{ -0.40}$ \\ 
SDSS J101614.26+520915.7 & $<0.10 $\\ 
SDSS J101954.54+082515.0 & 20.00$^{+8.30 }_{ -10.00 }$ \\ 
SDSS J104433.04-012502.2 & $<0.10 $\\ 
SDSS J110637.16+522233.4 & 1.90$^{+1.00 }_{ -1.00 }$ \\ 
SDSS J110853.98+522337.9 & $<0.10 $\\ 
SDSS J111859.56+075606.5 & 5.20$^{+3.20 }_{ -4.00 }$ \\ 
SDSS J112300.25+052451.0 & 15.90$^{+2.00 }_{ -4.00 }$ \\ 
UM 425                & 1.20$^{+0.1 }_{ -0.2 }$ \\ 
SDSS J112432.14+385104.3 & 13.00$^{+5.70 }_{ -2.50 }$ \\ 
SDSS J113537.67+491323.2 & $<0.10 $\\ 
SDSS J114111.61-014306.6 & 2.50$^{+1.00 }_{ -0.60 }$ \\ 
SDSS J122637.02+013015.9 & 6.00$^{+1.90 }_{ -2.20 }$ \\ 
SDSS J122708.29+012638.4 & 1.40$^{+1.00 }_{ -0.80 }$ \\ 
SDSS J123637.45+615814.4 & 1.50$^{+0.40 }_{ -1.00 }$ \\ 
LBQS 1235+1807B & $<0.10 $\\ 
SDSS J124559.59+570053.1 & $<0.50 $ \\ 
SDSS J132401.53+032020.6 & 12.70$^{+6.40 }_{ -4.00 }$ \\ 
SDSS J132827.07+581836.9 & 29.00$^{+12.00 }_{ -9.00 }$ \\ 
SDSS J133639.40+514605.2 &  $<0.10 $\\ 
SDSS J134059.24-001944.9 & $<0.10 $\\ 
HELLAS2XMM J140049.1+025850 & 11.00$^{+5.00 }_{ -5.00 }$ \\ 
SDSS J142555.22+373900.7 & $<0.10 $\\ 
SDSS J142652.94+375359.9 & 0.95$^{+0.50 }_{ -0.30 }$ \\ 
SDSS J144625.48+025548.6 & 7.60$^{+3.50 }_{ -4.50 }$ \\ 
SDSS J150858.15+565226.5 & 3.20$^{+0.45 }_{ -1.00}$ \\ 
SDSS J151729.70+001652.6 & $<0.10 $\\ 
SDSS J153322.80+324351.4 & 0.90$^{+1.20 }_{ -0.30 }$ \\ 
SDSS J203941.04-010201.6 & 3.00$^{+2.00 }_{ -1.50 }$ \\ 
SDSS J231850.79+002552.6 & 8.30$^{+5.00 }_{- 4.00 }$ \\

\hline 
\end{tabular}\\ 
\begin{flushleft} 
\small{Notes: Col.(1): Source name;  Col.(2): Neutral hydrogen column density estimated from hardness ratio. Errors correspond to 1$\sigma$ confidence intervals and estimated from the count-rate errors.} 
\end{flushleft} 
\end{table*}

%\end{appendix} 
%\\clearpage 
%\\begin{appendix} 
%\\section{Source Spectra}\label{appendix:1} 
 
%\\begin{figure*} 
%\\centering 
%\\subfigure{ 
%%\ label for subfigure 
%\\includegraphics[width=8cm]{0043p0046.ps}\hspace{2cm} 
%\\includegraphics[width=8cm]{0043p0052.ps}}  

\end{document}